\documentclass[11pt,a4paper]{article}
\usepackage[margin=2.5cm]{geometry}
\usepackage[varg]{txfonts}
\usepackage{graphicx}
\usepackage{booktabs}
\usepackage{authblk}
\usepackage{etoolbox}
\usepackage[hidelinks]{hyperref}
\apptocmd{\thebibliography}{\setlength{\itemsep}{0pt}\setlength{\parskip}{0pt}}{}{}

\begin{document}

\title{A study of CMS analysis pipelines through the Integration Challenge\thanks{Prepared for the proceedings of the \href{https://indico.cern.ch/event/1471803/contributions/6968224/}{presentation of the same title} at CHEP 2026.}}
\author[1]{Mohamed Aly\thanks{e-mail: mohamed.aly@cern.ch}}
\author[1]{Peter Elmer}
\author[1]{Peter Fackeldey}
\author[2]{Oksana Shadura}
\affil[1]{Princeton University, Princeton, NJ 08544, USA}
\affil[2]{University of Nebraska--Lincoln, Lincoln, NE 68588, USA}
\date{}

\maketitle

\begin{abstract}
The upcoming High-Luminosity Large Hadron Collider (HL-LHC) at CERN will deliver an unprecedented volume of data for High Energy Physics (HEP).
This wealth of information offers significant opportunities for scientific discovery, but its scale challenges traditional analysis workflows.
In this paper, we present CMS analysis pipelines being developed to meet HL-LHC demands.
These pipelines build on the broader scientific Python ecosystem, complemented by solutions specifically designed for HEP.
A central focus of this work is the Integration Challenge, an IRIS-HEP led effort aimed at assessing the readiness of developed software stack to be used in real world physics analysis and improving the readiness of analysis facilities for the HL-LHC era.
The Integration Challenge acts as an end-to-end integration test: by implementing a complete physics analysis pipeline, it evaluates tool interoperability and the overall user experience for analysts.
The current pipeline includes columnar data processing, machine learning, statistical inference, and visualization tasks covering a variety of CMS analysis scenarios.
In addition, the Integration Challenge explores efficient strategies for delivering skimmed data using diverse tools and data formats, as well as evaluating the ServiceX data-delivery system for HEP analyses.
Throughout the testing phase, we also investigated several prototype services, such as histogram-as-a-service capabilities, along with other emerging services that may support future HL-LHC analysis workflows.
\end{abstract}

%%%%%%%%%%%%%%%%%%%%%%%%%%%%%%%%%%%%%%%%%%%%%%%%%%%%%%%%%%%%%%%%%%%%%%%%%%%%%%
\section{Introduction}
\label{sec:intro}

The High-Luminosity Large Hadron Collider (HL-LHC)~\cite{hllhc} will deliver considerably more data than the experiments process today.
Current analysis workflows, with turnaround times of hours or days, do not scale to datasets an order of magnitude larger.

In CMS, most physics analyses start from centrally produced data tiers.
The NanoAOD tier~\cite{nanoaod} is about 1~kB per event, less than 1\% of an event in the RECO tier, the full output of event reconstruction, and is the input for more than half of CMS analyses.
A typical user workflow produces custom ntuples or NanoAODs from the central samples, skims them (discarding events and branches the analysis does not need), applies selections, corrections and systematic variations, fills histograms, and ends with a statistical analysis of the histograms.

To prepare for the HL-LHC, IRIS-HEP~\cite{irishep} and the experiments run community challenges.
A challenge implements a target workload at a defined scale, runs it on production infrastructure, and uses the outcome to test readiness and expose bottlenecks in the high-energy physics (HEP) software stack.
The Analysis Grand Challenge (AGC)~\cite{agc-chep} defined a representative physics analysis task on open data, small in scale but complete in its workflow aspects.
The 200~Gbps challenge~\cite{gbps200} focused on pure data throughput at an HL-LHC-relevant scale, with an intentionally minimal computation.
The Integration Challenge (IC) combines the two and adds realism: a complete physics analysis, implemented end to end with the tools analysts actually use, run at scale on user-facing analysis facilities\footnote{Shared, interactive computing services aimed at end-user analysis; the Analysis Facilities White Paper~\cite{afwhitepaper} defines the concept.}.
The IC exists in an ATLAS flavour~\cite{held-atlas-ic} and the CMS flavour described here.
Its aims are to stress-test software and infrastructure with a realistic workload, to research alternative workflows, and to produce concrete recommendations to users, to CMS, and to software developers.

The statistical analysis stage is work in progress and is not covered here.
Complementary contributions at this conference cover the facility perspective~\cite{shadura-ic}, the ATLAS Integration Challenge~\cite{held-atlas-ic}, and the services and tools studied here~\cite{fackeldey-haas,galewsky-sx,neumeister-purdue,krommydas-coffea,hageboeck-root7}.

%%%%%%%%%%%%%%%%%%%%%%%%%%%%%%%%%%%%%%%%%%%%%%%%%%%%%%%%%%%%%%%%%%%%%%%%%%%%%%
\section{The Integration Challenge}
\label{sec:setup}

\subsection{Analysis pipeline}
\label{sec:implementation}

\begin{figure}[t]
  \centering
  \includegraphics[width=0.76\textwidth]{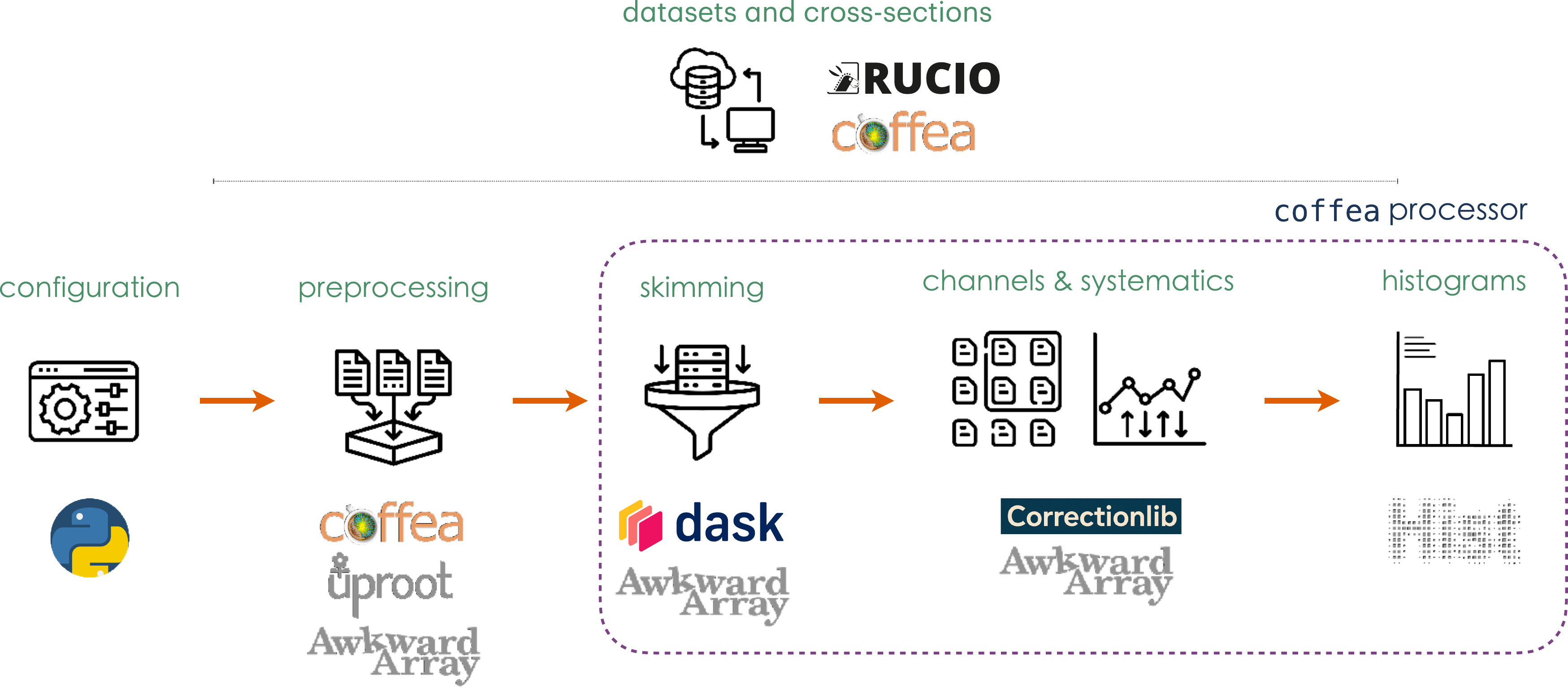}
  \caption{The Integration Challenge analysis pipeline.
  The coffea processor encompasses skimming, channel and systematics evaluation, and histogramming; datasets and cross-sections are resolved through Rucio and coffea.}
  \label{fig:pipeline}
\end{figure}

The IC analysis is a distributed, columnar pipeline built on the scientific Python ecosystem, with HEP-specific components from the Scikit-HEP~\cite{scikithep} and IRIS-HEP software stacks.
The implementation is publicly available~\cite{ic-code}.
Figure~\ref{fig:pipeline} shows the pipeline.
Datasets and cross-sections are resolved via \texttt{Rucio}~\cite{rucio} through \texttt{coffea}~\cite{coffea}.
A Python configuration drives all stages.
Preprocessing builds the fileset metadata with \texttt{coffea}, \texttt{uproot}~\cite{uproot} and \texttt{Awkward Array}~\cite{awkward}: the file lists, the per-file event counts, and the boundaries of the work items.
Work items (chunks) are defined in events rather than files: each file is split into chunks of a configurable number of events, and one chunk is one task on the cluster.

The core of the analysis is a \texttt{coffea} processor, executed on distributed workers with \texttt{Dask}~\cite{dask}.
It applies event selections, evaluates corrections and systematic variations with \texttt{correctionlib}~\cite{correctionlib} and \texttt{Awkward Array}, and fills histograms with \texttt{hist}~\cite{hist}.
Each systematic variation re-evaluates the affected parts of the event processing, so the variations multiply the computation per chunk.

Skimming runs inside the processor by default: each processed chunk can be written back to storage as a skimmed file, so one chunk produces one output file.
The performance consequences of this choice appear in Section~\ref{sec:skimming}.
A standalone skimming step outside the processor is also supported, and the \texttt{ServiceX} pre-skimming of Section~\ref{sec:servicex} uses this route.

\subsection{The physics analysis and its inputs}
\label{sec:inputs}

The analysis is a search for a hypothetical heavy resonance $Z'$ decaying to a top quark pair, following the CMS Run~2 analysis~\cite{zprime}.
It uses the single-lepton final state, where one top quark decays hadronically and the other leptonically, restricted to the muon channel.
It was chosen because it is straightforward as a physics analysis and is not tailored for benchmarking.
A representative subset of the systematic uncertainties is implemented, amounting to 66 variations of the event processing.

The inputs are Run~2 CMS NanoAOD samples for data and all relevant simulated processes across the 2016--2018 data-taking years: 10.8~TB in 13\,520 files, $6.83\times10^{9}$ events, half of it $t\bar{t}$ simulation and data.
The file population is strongly non-uniform.
The average file holds $5.05\times10^{5}$ events and the median $1.53\times10^{5}$, data and $t\bar{t}$ files hold $\mathcal{O}(10^{6})$ events each, whereas a large number of small $W$+jets files hold $\mathcal{O}(10^{5})$ each.
This spread matters for scheduling, because work items are defined in events rather than files.
Its consequences appear throughout Section~\ref{sec:nominal}.
Two branch selections are used throughout the measurements.
The narrow selection is what the $Z' \to t\bar{t}$ analysis actually requires: 1.5\% of the branches\footnote{With the implemented subset of systematic uncertainties. The complete set of the original analysis would raise this to roughly 3--4\%.}, about 5\% of the stored bytes (roughly 600~GB), at a mean compression ratio of 0.23.
The wide selection is the branch list of the 200~Gbps challenge and covers roughly a quarter of the stored bytes.
Materialising a selection means forcing those branches to be read and decompressed into memory, whether or not the analysis uses them.

\subsection{The Coffea-Casa analysis facility}
\label{sec:facility}

All measurements were performed on the Coffea-Casa analysis facility at the University of Nebraska--Lincoln~\cite{coffea-casa,albin-composable}, which provides interactive JupyterLab entry points, token-based read access to CMS data, and S3 object storage.
Central to this exercise, \texttt{XCache} servers~\cite{xcache}, deployed and tuned during the 200~Gbps challenge~\cite{albin-tuning}, cache CMS data in front of the CMS data federation.
All measurements use the facility's fast-turnaround Kubernetes queue with 800 single-core workers, where \texttt{Dask} clusters are provisioned on demand through \texttt{Dask Gateway} and worker pods are co-located with the \texttt{XCache} pods, so traffic remains inside the cluster~\cite{shadura-ic}.

%%%%%%%%%%%%%%%%%%%%%%%%%%%%%%%%%%%%%%%%%%%%%%%%%%%%%%%%%%%%%%%%%%%%%%%%%%%%%%
\section{Measurements of the nominal configuration}
\label{sec:nominal}

The nominal configuration approximates what a CMS analyst would use today: NanoAOD stored as TTrees, ROOT's established columnar event format, compressed with LZMA, 500k-event chunks, the narrow branch selection, and the 66 systematic variations evaluated on the fly.
Four metrics are quoted: data throughput (bytes delivered to the processors per unit time), CPU event rate (events divided by total CPU time), total event rate per core (events divided by wall time and core count), and wall time.
Table~\ref{tab:overview} collects the throughput measurements.

As a reference ceiling (last row of Table~\ref{tab:overview}), a distributed copy of the entire input over the network into the facility caches, across the 800 workers, sustained 173~Gbps on average and completed in seven minutes.

\begin{table}[t]
  \centering
  \small
  \scalebox{0.97}{%
  \begin{tabular}{llcccc}
    \toprule
    Implementation & Branches & Chunk & Rate & CPU ev.\ rate & Wall time \\
     & & [$10^3$] & [Gbps] & [kHz] & \\
    \midrule
    Read-only, plain \texttt{uproot} & wide & 200 & 43 & $\sim$20 & $\sim$7 min \\
    Read-only, \texttt{coffea} & wide & 200 & 42 & $\sim$17 & $\sim$14 min \\
    Read-only, \texttt{coffea} & wide & 500 & 50 & $\sim$15 & $\sim$13 min \\
    Full IC, no systematics & narrow & 500 & 18 & 85.8 & $\sim$5 min \\
    Full IC, no systematics & narrow+wide & 500 & 40 & 21.2 & $\sim$10 min \\
    Full IC, 66 variations & narrow & 500 & 6 & 12.6 & $\sim$20 min \\
    Distributed copy to cache & -- & -- & 173 & -- & $\sim$7 min \\
    \bottomrule
  \end{tabular}}
  \caption{Throughput measurements over identical input files on 800 cores, changing one variable at a time.}
  \label{tab:overview}
\end{table}

\subsection{Read-only throughput and chunk-size dependence}
\label{sec:gbps}

The 200~Gbps challenge~\cite{gbps200} demonstrated high-throughput columnar reading with a read-only workload: the requested branches are decompressed into arrays in memory and nothing further is computed.
This workload is repeated here to confirm that the facility still reproduces the challenge results, so that comparisons to them are free of infrastructure differences.
Two implementations are used, differing only in the reading layer.
The plain \texttt{uproot} implementation is the original 200~Gbps challenge code: it opens each file once, in a single task, and reads the requested branches in steps of a fixed number of events.
The \texttt{coffea}-based reader of the IC pipeline splits each file into chunks that are independent tasks: the same file can be opened several times, each task carries its own setup and network communication, and at least one auxiliary branch is materialised alongside each requested one.
A granularity of 200k events therefore means a step size within one file open for \texttt{uproot}, and a chunk size across tasks for \texttt{coffea}.
Both implementations reproduce the challenge results closely (first three rows of Table~\ref{tab:overview}): 43.1~Gbps against 42.1~Gbps at the same 200k granularity, a small and well-understood gap that follows directly from these differences.
The \texttt{coffea} runs read 3.94~TB of compressed data from the wide selection, while the plain \texttt{uproot} run read 2.39~TB: 1060 files failed to process and are excluded from its measurement.

Chunk size is a single global parameter with a direct trade-off.
Raising it from 200k to 500k events increases the read-only throughput from 42.1 to 50.3~Gbps: fewer tasks mean less per-chunk overhead and fewer repeated file opens.
The cost is worker memory, which grows with the number of events held in flight.
This trade-off motivates work on dynamic chunk sizing and on in-memory compression~\cite{fackeldey-compression}, which would allow larger chunks without a proportional rise in worker memory.

\subsection{Nominal analysis performance}
\label{sec:baseline}

The fourth and sixth rows of Table~\ref{tab:overview} give the nominal analysis without and with the systematics loop.
Table~\ref{tab:baseline} adds the resource profile of the two runs.
Without systematics the analysis processes the full inputs in about five minutes at 17.7~Gbps.
The 66 variations multiply the computation per chunk, and the run time increases to about 20 minutes at 5.8~Gbps, with a long tail in the chunk-runtime distribution.
Worker CPU utilisation is essentially 100\% in both configurations: the pipeline is compute-bound, and the systematics evaluation is the dominant workload.

\begin{table}[t]
  \centering
  \small
  \scalebox{0.95}{%
  \begin{tabular}{lcc}
    \toprule
    Metric & No systematics & Systematics \\
    \midrule
    Total event rate per core [kHz/core] & 28.9 & 9.5 \\
    Mean chunk runtime [s] & 3.7 & 25.1 \\
    Median chunk runtime [s] & 3.9 & 15.8 \\
    Mean worker memory [GB] & 0.9 & 1.2 \\
    Peak worker memory [GB] & 1.3 & 2.0 \\
    CPU utilisation & 100\% & 100\% \\
    \bottomrule
  \end{tabular}}
  \caption{Resource profile of the nominal analysis on 800 cores at 500k-event chunk size, without and with the systematics loop (66 variations).}
  \label{tab:baseline}
\end{table}

\begin{figure}[t]
  \centering
  % trim order: left bottom right top (values to be set on Overleaf)
  \includegraphics[width=0.47\textwidth, trim={6 0 0 6}, clip]{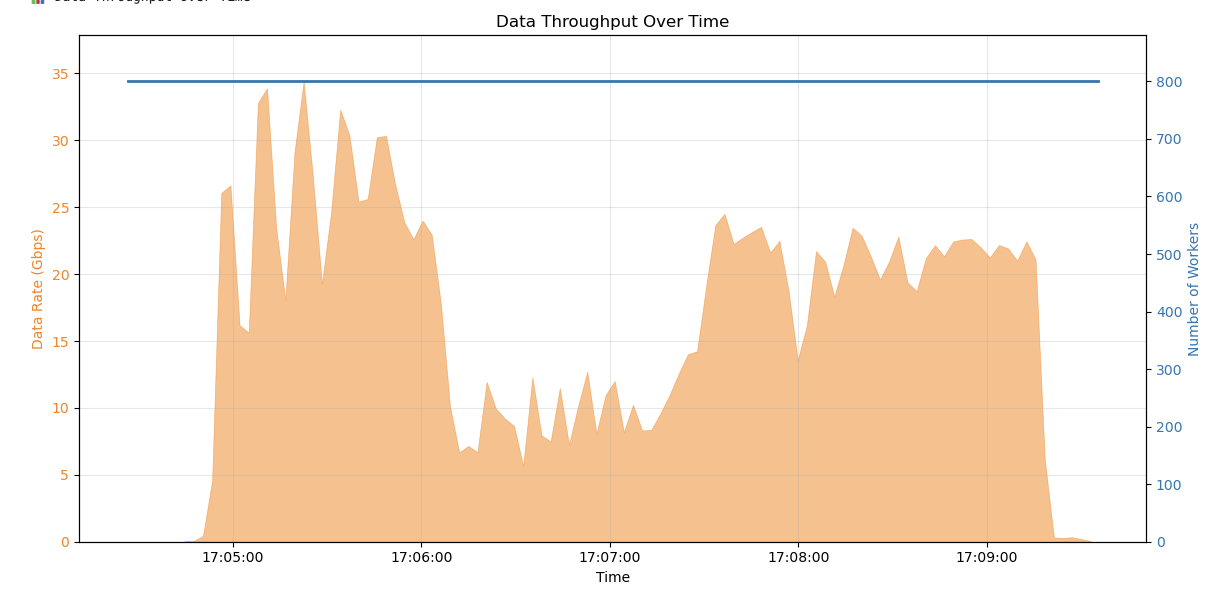}
  \includegraphics[width=0.51\textwidth]{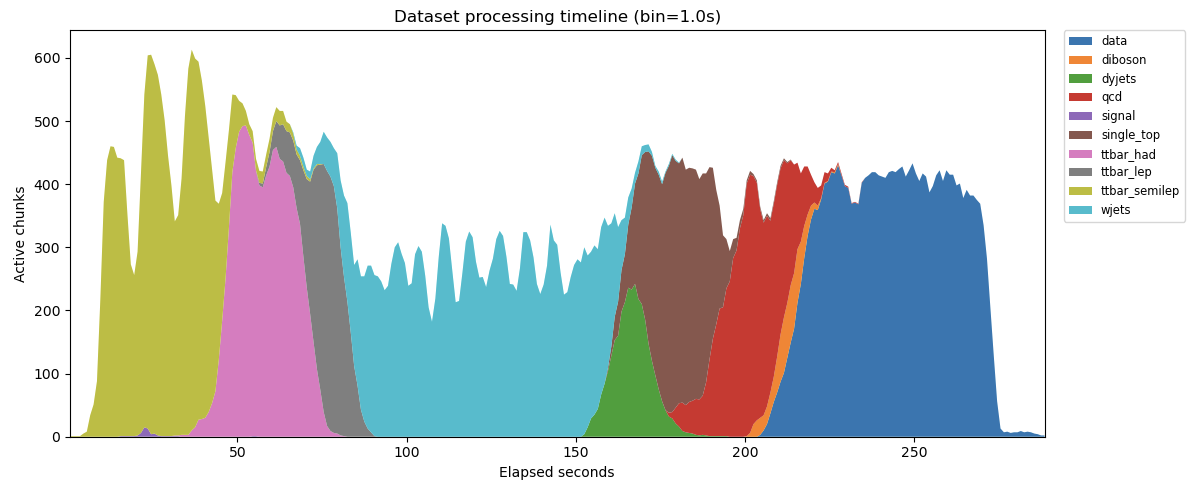}
  \caption{Data throughput over time (left) and dataset processing timeline (right) for the nominal analysis without systematic variations.
  The long central stretch of reduced throughput is dominated by the many small $W$+jets files.}
  \label{fig:timelines}
\end{figure}

The dataset processing timeline of the run without systematics and its throughput over time (Figure~\ref{fig:timelines}) show how the processing time is distributed.
The physics processes are handled in distinct, largely non-overlapping phases.
Large-file datasets ($t\bar{t}$, data) process at high throughput.
The $W$+jets datasets, spread over many small files, occupy the cluster for a long stretch at visibly reduced throughput.
Small files produce undersized work items dominated by per-chunk overhead.
Large files produce long chunks that finish late in each phase, since per-chunk runtime scales with the events per chunk.
Infrastructure for merging input files would directly improve this.
Memory use remains modest (Table~\ref{tab:baseline}), but the systematics loop adds steady growth over the run, which further motivates smarter chunking and memory-reduction techniques such as in-memory compression~\cite{fackeldey-compression}.

The nominal analysis reads only the narrow branch selection: 1.5\% of the branches and about 5\% of the stored bytes.
To test whether this limits the throughput, the wide selection was materialised on top of the nominal narrow one, increasing the data read by a factor of five with no other change.
Throughput rose to 40~Gbps, 2.5 times the nominal value (Table~\ref{tab:overview}).
The CPU event rates support the same conclusion: the read-only workload on the wide selection reaches only about 20~kHz, while the full pipeline on the narrow selection reaches 85.8~kHz, so decompressing a quarter of the branches costs considerably more CPU than analysing four percent of them.
Hence, the nominal analysis is limited by the small data volume read per task, not by the infrastructure.
%%%%%%%%%%%%%%%%%%%%%%%%%%%%%%%%%%%%%%%%%%%%%%%%%%%%%%%%%%%%%%%%%%%%%%%%%%%%%%
\section{Workflow variations}
\label{sec:research}

\subsection{Storage format: TTree and RNTuple}
\label{sec:formats}

The NanoAOD baseline is TTree with LZMA compression.
RNTuple, the columnar successor of TTree in ROOT~7, pairs naturally with zstd compression and is the future ROOT default~\cite{rntuple,hageboeck-root7}.
The IC inputs were converted to RNTuple, with a production that has not yet been optimised, and the nominal analysis without systematics was repeated.
Wall time is unchanged at about five minutes, with 15.2~Gbps against 17.7~Gbps for TTrees and an essentially unchanged event rate per core (26.1 against 28.9~kHz).
The difference is CPU efficiency: 281~kHz CPU event rate for RNTuple against 85.8~kHz for TTree, more than three times higher.
RNTuples spend much less CPU time than TTrees, but at the low read fractions of this analysis enough non-CPU time, dominated by network I/O and task scheduling waits, remains to cancel the gain.
Format and compression matter in proportion to how much data is actually read.
Larger reads per task would make the RNTuple advantage visible in wall time as well.

\subsection{Histogram aggregation: tree reduction and histogramming-as-a-service}
\label{sec:haas}

In the baseline setup, every task fills its own copy of the output histograms, and the copies are summed pairwise up a reduction tree until one aggregated histogram reaches the user.
Partial sums occupy worker memory for the whole run, so memory use climbs steadily and peaks near the end, exactly where a worker failure discards the largest partial results.
No result exists until the tree completes.
Histogramming-as-a-service (HaaS)~\cite{fackeldey-haas}, prototyped in \texttt{histserv}~\cite{histserv}, instead sends each fill to a dedicated histogram server, which aggregates centrally and serves the result together with the raw metrics.
No reduction tree is needed, and the histogram exists on the server even after partial processing.
With HaaS, the full analysis with systematics uses more than 20\% less worker memory and shows no memory growth over the run, at minimal implementation overhead.
Since memory limits the usable chunk sizes (Section~\ref{sec:gbps}), this is a promising direction for reducing time-to-insight.

\subsection{Skimming}
\label{sec:skimming}

Skimming addresses the read-fraction problem directly: much of the nominal run is spent reading data that is discarded immediately, and a skim reduces the downstream input size.
In the IC, skimming runs inside the \texttt{coffea} processor with per-chunk saves and writes TTree, Parquet, or RNTuple outputs.
Saving to disk is optional, and a second analysis pass then runs on the skimmed fileset.
Because one chunk produces one output file, $M$ chunks produce $M$ skimmed files.
Outputs go to \texttt{XRootD}-managed storage (TTree, RNTuple, X.509 authentication) or S3 object stores (Parquet, key-based authentication), via direct remote I/O, a subprocess copy through local scratch, or \texttt{fsspec} caching.

With systematics enabled, skimming the full 10.8~TB input takes about 300~s for either output format, and the analysis pass on the skims reads only around 50~GB, two orders of magnitude less, from a simple event-level cut plus branch selection.
The analysis pass decreases from 857~s without skims to 668~s on TTree skims, and from 874~s to 703~s for RNTuples.
The wall-time gain is smaller than the volume reduction suggests: the per-chunk saves produce a very large number of small files, which severely limits read throughput and, on the S3 path, encounters object-store rate limits.
A file-merging step is required to realise the full benefit of skimming.
ServiceX is a candidate to provide one as a service.

\subsection{ServiceX data delivery and pre-skimming}
\label{sec:servicex}

\texttt{ServiceX}~\cite{servicex,galewsky-sx} is a data delivery service for HEP.
Transformer pods filter events and columns on request and deliver the reduced output, typically to an object store.
The transformers run at the site that hosts the input data, so the expensive wide read is performed once, without crossing the wide-area network.
In the IC workflow, \texttt{ServiceX} acts as a pre-skimmer and delivers a skimmed fileset directly into the \texttt{coffea} processor.

This variation was measured at the Purdue analysis facility~\cite{neumeister-purdue}, comparing four data-access configurations (Table~\ref{tab:purdue}): (1) remote \texttt{XRootD} reads from the Fermi National Accelerator Laboratory (FNAL) over the wide-area network; (2) \texttt{XRootD} reads from local storage at Purdue; (3) reads through a local \texttt{XCache} in front of that storage; and (4) \texttt{ServiceX} pre-skimming, with the skimmed output written to the local S3 object store.
Moving from remote reads (1) to a local \texttt{XCache} (3) raises the read rate five-fold and turns the workload from I/O-bound to mostly CPU-bound.
With \texttt{ServiceX} pre-skimming (4) the read rate drops again and the workload returns to being almost entirely I/O-bound, because the uncompressed output shifts the bottleneck to the S3 store.
Local caching combined with \texttt{ServiceX} delivery improves throughput substantially, provided the output compression issue is resolved.
Work on this is ongoing.

\begin{table}[t]
  \centering
  \small
  \scalebox{0.90}{%
  \begin{tabular}{lcccc}
    \toprule
     & (1) XRootD & (2) XRootD & (3) XCache & (4) ServiceX \\
     & (FNAL) & (Purdue) & & skimming \\
    \midrule
    Preprocessing time & 3 min & 2 min & 30 s & 10 min \\
    Processing time & 18 min & 11 min & 4 min & 4.5 min \\
    Processing data rate [Gbps] & 4.5 & 7.5 & 23 & 17 \\
    CPU bottleneck fraction & 10\% & 24\% & 71\% & 6\% \\
    I/O and waiting fraction & 90\% & 76\% & 29\% & 94\% \\
    \bottomrule
  \end{tabular}}
  \caption{IC measurements at the Purdue analysis facility for the four data access configurations, from Ref.~\cite{neumeister-purdue}.
  The first column uses 1000 single-core workers, the others 500 two-core workers.}
  \label{tab:purdue}
\end{table}

\subsection{Scale tests of coffea developments}
\label{sec:coffea-dev}

The IC also serves as a scale test for new \texttt{coffea} features~\cite{krommydas-coffea}; the measurements above use \texttt{coffea} 2026.4/5.0.
Virtual arrays, the lazy NanoEvents backend where branches are loaded on first access, ran reliably at scale throughout.
Tracing with preloading, which dry-runs the processor to discover the accessed branches and bundles their network requests into a single fetch, showed no significant impact for this analysis, which reads too few branches to benefit, in contrast to its ATLAS counterpart~\cite{held-atlas-ic}.
Parquet reading through the same \texttt{coffea} executor as ROOT inputs proved important for the skimming studies.

%%%%%%%%%%%%%%%%%%%%%%%%%%%%%%%%%%%%%%%%%%%%%%%%%%%%%%%%%%%%%%%%%%%%%%%%%%%%%%
\section{Conclusions and outlook}
\label{sec:conclusions}

The Integration Challenge runs a realistic CMS analysis end to end on analysis facilities with the HEP Python stack, and uses it to measure where HL-LHC era workflows stand.
The main findings are the following.
Analyses that use small fractions of their input data are throughput-inefficient; the nominal analysis reads 1.5\% of the branches and is limited precisely by this small read fraction.
Small files degrade performance, and file-merging infrastructure, for inputs and skim outputs alike, would be broadly useful.
Chunk size trades performance against worker memory, motivating smart chunking and memory reduction such as in-memory compression.
RNTuples are considerably more CPU-efficient than TTrees, but the benefit only becomes visible once substantial data volumes are read.
Histogramming-as-a-service reduces worker memory by more than 20\% at minimal overhead.
\texttt{ServiceX} delivery with local caching improves runtime provided its output is compressed.
The new \texttt{coffea} releases with virtual arrays performed reliably at the full scale of the exercise.

The next steps of the IC are the remaining systematic variations, which will raise the read volumes towards more realistic levels, the integration of the statistical analysis stage, and continued work with the ecosystem on file merging, dynamic chunking, and memory reduction.

\section*{Acknowledgements}

This work was supported by the National Science Foundation under Cooperative Agreements OAC-1836650 and PHY-2323298.

%%%%%%%%%%%%%%%%%%%%%%%%%%%%%%%%%%%%%%%%%%%%%%%%%%%%%%%%%%%%%%%%%%%%%%%%%%%%%%
\bibliographystyle{woc}
\bibliography{refs}

\end{document}